\documentclass[useAMS,usenatbib]{mn2e}
\usepackage{graphicx}
\begin{document}

\title[Density and Width of Saturated Ly$\alpha$ Absorption]
{Correlation between the Mean Matter Density and the Width 
	of the Saturated Ly$\bmath\alpha$ Absorption}	

\author[Hu Zhan]{Hu Zhan\thanks{Email: zhanhu@physics.arizona.edu} \\
Department of Physics, University of Arizona, Tucson, AZ 85721, USA}

\maketitle

\begin{abstract}
We report a scaling of the mean matter density with the width of 
the saturated Ly$\alpha$ absorptions. This property is established 
using the ``pseudo-hydro'' technique \citep{c98}. It provides 
a constraint for the inversion of the Ly$\alpha$ forest, 
which encounters difficulty in the saturated region. With a Gaussian 
density profile and the scaling relation, a simple inversion of the 
simulated Ly$\alpha$ forests shows that the one-dimensional mass power 
spectrum is well recovered on scales above 2 \mbox{$h^{-1}$Mpc}, or roughly
$k\la 0.03$ \mbox{km$^{-1}$s}, at $z=3$. The recovery underestimates the 
power on small scales, but improvement is possible with a more 
sophisticated algorithm. 
\end{abstract}


\section{Introduction} \label{sec:intro}
The Ly$\alpha$ forest has offered a unique way to study the large-scale 
structure of the universe at high redshifts. It is based on the theory 
that the Ly$\alpha$ forest is a result of the absorptions of quasar
continuum by the diffusely distributed and photoionized intergalactic 
medium (IGM), and the IGM traces the
density of the underlying mass field on scales larger than the Jeans 
length \citep{b93,c94,bgf95,pmk95,v02}. In other words, the baryon 
density in Ly$\alpha$ absorption systems is roughly proportional 
to the matter density, when $\rho\la 10$ \citep{bd97,gh98,z98}.

To extract the baryon density $\rho_{\mathrm{b}}$ or the mass density 
$\rho$, one may invert 
the transmitted flux $F$ of the Ly$\alpha$ forest using
\begin{equation} \label{eq:flux-dens}
F = e^{-\tau} \simeq e^{-A\rho_\mathrm{b}^\beta} \simeq e^{-A\rho^\beta},
\end{equation}
where: $\tau$ is the Ly$\alpha$ optical depth;
$A$ depends on the temperature of IGM,
the \mbox{H\,{\sc ii}} photoionization rate, and redshift \citep{r97,c98}; 
and $\beta$, related to equation of state of the IGM \citep{hg97}, is in 
the range 1.6--1.8 \citep{c99}. The densities are in units of 
their corresponding cosmic mean density, and so the overdensity 
$\delta = \rho - 1$.

When the density 
is high enough, the spectrum is saturated, i.e. $F \simeq 0$. With noises
and uncertainties in the spectrum, the direct inversion using equation 
(\ref{eq:flux-dens}) is very unreliable in the saturated region. Despite 
the difficulty, methods of direct inversion are systematically developed,
for example, with Lucy's method by \citet{nh99}, and with Bayesian method 
for a three-dimensional inversion by \citet{pvr01}. One may also use 
higher order lines to recover the optical depth and the underlying density
\citep{cs98,ast02}, even though the contamination by lower order lines needs
to be carefully removed. Once the cosmic density field is obtained, many 
statistics, such as the mass power spectrum, can be measured.

Another approach is taken by \citet{c98,c99,c02} to map the mass power
spectrum directly from the flux power spectrum of the Ly$\alpha$ forest 
without an inversion. Thus the saturation problem is avoided. 
However, a close examination of the Fourier transform of equation 
(\ref{eq:flux-dens}) shows that powers on different scales are mixed 
by the non-linear density--flux relation (see Section \ref{sec:map}). 
The mixing depends on the underlying density field, 
and it is hard to predict analytically.

If the inversion is necessary, a proper treatment in the saturated region
has to be developed. In many physical systems, sizes are often correlated
with other quantities such as masses and densities. For example, more 
massive stars or dark matter haloes have larger sizes, but 
lower mean densities \citep{bm98,nfw96}. One may expect a similar 
trend for the saturated Ly$\alpha$ absorption.
On the contrary, the mean density is found to increase with the 
width of saturation. This is due to the fact that the IGM is very 
diffuse and far from virialization, while the other objects mentioned are 
the opposite. 

This paper is organised as follows. Section \ref{sec:scaling} presents the 
scaling relation. A simple inversion with the scaling is tested in Section 
\ref{sec:test} for both the one-dimensional mass power spectrum and the 
one-point distribution function. Section \ref{sec:map} examines the mapping 
between the mass power spectrum and the flux power spectrum.
The discussion and conclusions are given in Section \ref{sec:conclusion}. 
Unless addressed otherwise, the power spectra below are implicitly
one-dimensional.

\section{The Scaling}  \label{sec:scaling}
Neglecting the probability that two physically separate absorption systems 
fall in the same redshift coordinate, one can associate a saturated 
absorption in the Ly$\alpha$ forest with a single high density region. 
If we assume further that there is no substructure present, and the IGM 
evolves more or less the same way everywhere, then the size of the saturated 
region has to be tightly correlated with its mean density \citep{s01}. In 
reality, the neglected elements above and uncertainties elsewhere will 
introduce a spread to the correlation. 

\subsection{Simulated Ly$\alpha$ Forest} \label{sec:sim}
It is demonstrated by various hydrodynamical simulations \citep{bd97, gh98, 
z98, b99, dhk99, dco01} that the baryon density, the \mbox{H\,{\sc i}} 
column density, and the IGM temperature correlate 
well with the underlying dark matter density for $\rho \la 10$. This 
leads to the useful ``pseudo-hydro'' technique \citep{c98} for simulating
the Ly$\alpha$ forest with collisionless N-body simulations instead of 
hydrodynamical simulations. Basically it draws the line-of-sight (LOS) 
dark matter density in redshift space from N-body simulations, and then 
converts the density to flux using equation (\ref{eq:flux-dens}). 

\begin{figure}  
\includegraphics[width=84mm]{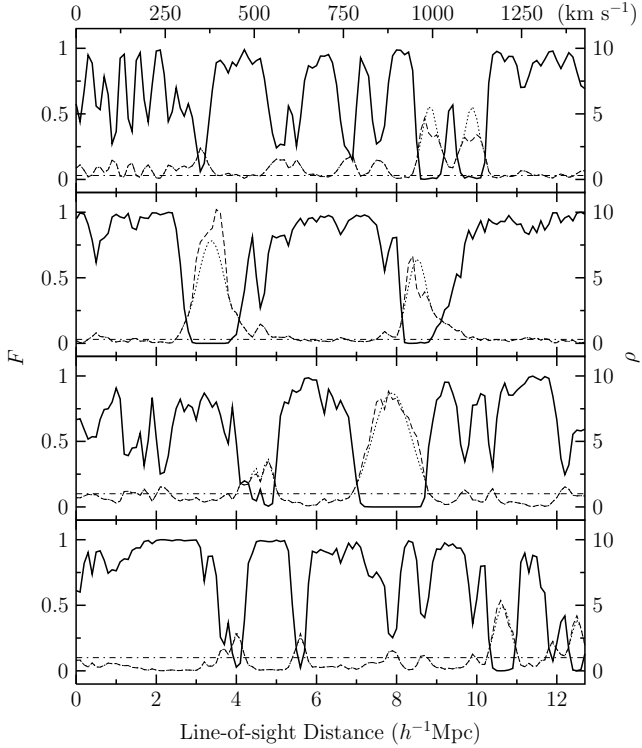}
\caption[f1]{Samples of the simulated Ly$\alpha$ forest from 
the LCDM model at $z=3$. The solid
lines are the flux $F$, the dashed lines are the matter density $\rho$, 
the dotted lines are the recovered density, and the horizontal dash-dotted
lines indicate the threshold flux level $\eta=0.03$ 
(see Section \ref{sec:test}) in the upper 2 panels, and $\eta =0.10$ in the 
lower 2 panels. The scale of the densities, in units of cosmic mean 
density, is indicated on the right axis. The flux and the density
are both presented in redshift space.}
\label{fig:sample}
\end{figure}

One may have concerns about the Jeans length, below which the pseudo-hydro 
technique may not be applicable. The comoving Jeans length $L_\mathrm{J}$ 
is
\begin{eqnarray} \label{eq:jeans}
L_\mathrm{J} &=& \frac{c_\mathrm{s} (1+z)}{\sqrt{G\rho}} =
\sqrt{\frac{8\pi \gamma k_\mathrm{B} T}{3\mu m_\mathrm{p} 
H_\mathrm{0}^2 \Omega \rho (1+z)}} 
\nonumber \\ 
&=& 440\ \mbox{$h^{-1}$kpc}\ T_\mathrm{4}^{1/2}[\Omega \rho (1+z)]^{-1/2},
\end{eqnarray}
where $c_\mathrm{s}$ is the speed of sound, $G$ is the gravitational constant, 
$\gamma=5/3$ is the ratio of specific heats, 
$k_\mathrm{B}$ is the Boltzman constant, $T$ is the temperature, $\mu=0.59$ 
is the mean molecular weight, $m_\mathrm{p}$ is the proton mass, $H_\mathrm{0} = 100$ 
\mbox{km s$^{-1}$ Mpc$^{-1}$}, $\Omega$ is the matter density in units 
of the critical density at present, and $T_\mathrm{4}\simeq 1.5$ is the temperature 
in units of $10^4$ K. It is understood that the first $\rho$ in equation
(\ref{eq:jeans}) is the actual density, while the rest are densities in
units of the cosmic mean density. For the low-density cold dark matter 
(LCDM) model at $z=3$ and 
$\rho=1$, $L_\mathrm{J}=490$ \mbox{$h^{-1}$kpc}. The Jeans length suggests that 
the smallest scale we can analyse reliably without hydrodynamical 
simulations is 0.5 \mbox{$h^{-1}$Mpc}. However, saturated regions usually
have $10 \ga \rho \ga 5$ (see Fig.~\ref{fig:sample}), and the temperature 
scales with the density as $T\propto \rho^{\alpha}$ with $\alpha$ = 
0.3--0.6 \citep{hg97}, so the Jeans length is reduced by a factor of 
1.4--2.2 in those regions. In addition, \citet{gh98} point out 
that the actual linear filtering scale is 1.5--2.5 times smaller than the
Jeans length at $z=3$, such that baryons trace the dark matter on scales 
even half of $L_\mathrm{J}$. Hence, the pseudo-hydro technique is sufficient for 
the analysis above the scale of 200 \mbox{$h^{-1}$kpc}.

A standard particle-particle-particle-mesh \citep[P$^3$M,][]{he81} 
code developed by \citet{jf94} is used to evolve
128$^3$ dark matter particles in a cubic box of 12.8 \mbox{$h^{-1}$Mpc} 
(comoving) each side. The initial power spectrum is given by the 
fitting formula from 
\citet{b86}. The model parameters are listed in Table \ref{tab:para}. All 
the models start from $z = 15$, and stop at $z = 3$ in 950 steps.
\begin{table}
\caption{Parameters of the N-body simulations.}
\label{tab:para}
\begin{tabular}{@{}cccccc}
\hline
Model & $\Omega$ & $\Lambda$ & $h$ & $\Gamma$ & $\sigma_\mathrm{8}$ \\
\hline
LCDM & 0.3 & 0.7 & 0.7 & 0.21 & 0.85 \\
OCDM & 0.3 & 0   & 0.7 & 0.21 & 0.85 \\
SCDM & 1.0 & 0   & 0.5 & 0.5  & 0.67 \\
TCDM & 1.0 & 0   & 0.5 & 0.25 & 0.60 \\
\hline
\end{tabular}

\medskip
{\em h} is the Hubble constant in units of 100 
\mbox{km s$^{-1}$ Mpc$^{-1}$}.\\
$\Gamma$ is the shape parameter in the power spectrum.\\
$\sigma_\mathrm{8}$ is the r.m.s. density fluctuation within a radius of 8 
\mbox{$h^{-1}$Mpc}.
\end{table}

The parameter $A$ in equation (\ref{eq:flux-dens}) is chosen to fit
the mean flux $\left< F(z) \right> = \exp[-\tau(z)]$, where 
$\tau(z)=0.0032(1+z)^{3.37}$ for \mbox{$z = 3$--$4$} \citep{k02}.
This mean flux formula is consistent with other observations 
\citep{l96, r97, m00}. We extrapolate the mean flux up to $z = 4.5$, 
which is not critical to the analysis in Section \ref{sec:rho-w}, 
but is nevertheless 
supported by simulations \citep{rpm98}. 
The constant $\beta$ in equation (\ref{eq:flux-dens}) is set to
$1.6$. Four samples of the Ly$\alpha$ forest are shown in 
Fig.~\ref{fig:sample}. They are drawn from the LCDM 
simulation at $z=3$. The corresponding LOS densities are plotted along
with the fluxes. It is evident that most of the Ly$\alpha$ lines arise 
where $\rho \la 10$.

\subsection{The Scaling} \label{sec:rho-w}
We define the width $w$ of the saturation as the distance between the two 
points that bracket the absorption at a given threshold flux level $\eta$.
In other words, it is the width of a region in which $F<\eta$. 
Since the flux $F$ in the saturated region is dominated by noise, one 
should set the threshold above the noise level. Within a reasonable 
range of spectral quality, we choose the 
threshold $\eta=0.03$ and $0.10$ for the tests below.

\begin{figure}  
\includegraphics[width=70mm]{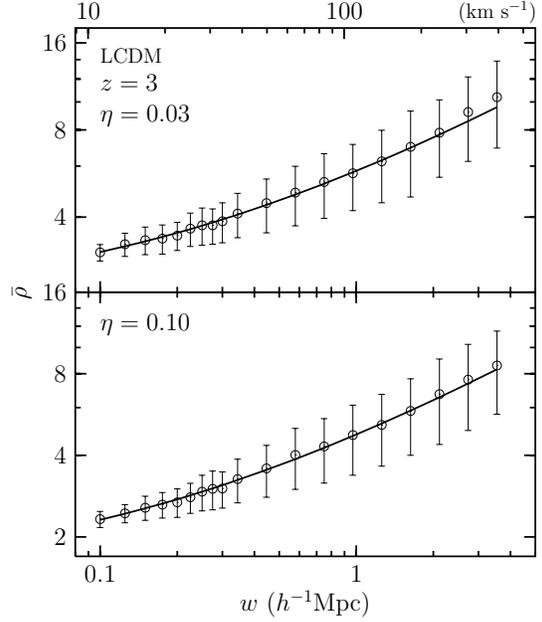}
\caption[f2]{The correlation between the mean density 
$\bar{\rho}$ and the width $w$ of the saturated region in the 
Ly$\alpha$ forest. The trend lines fit the data well beyond 
$2\ h^{-1}$Mpc, even though they are obtained by fitting only the data 
with $w \leq 2\ h^{-1}$Mpc. The data above $0.3\ h^{-1}$Mpc are binned in 
logarithmic intervals for readability. Similar treatment applies
to Figs \ref{fig:powerspect}, \ref{fig:powerspect-ost}, 
\ref{fig:powerspect-moc}, and \ref{fig:bdk} as well. 
\label{fig:scaling}}
\end{figure}

Once the width is determined, the mean density $\bar{\rho}$ can be
readily calculated from the corresponding LOS density field. Since the 
density and the flux are assigned only on grid points, interpolation is 
needed to find $w$ and $\bar{\rho}$. Fig.~\ref{fig:scaling} shows the
scaling relation for the LCDM model at $z=3$. It clearly
demonstrates that $\bar{\rho}$ increases with $w$. Furthermore, the 
correlation is reasonably tight for narrow saturations.
Notice that the trend lines are obtained by fitting only the 
data with $w \leq 2\ h^{-1}$Mpc, because wider saturations are very rare
as compared to the rest.

The $\bar{\rho}$--$w$ relation is well fitted by
\begin{equation}
\bar{\rho}=\rho_\mathrm{0} + a\ (w/h^{-1}\mbox{Mpc})^b.
\end{equation}
The term $\rho_\mathrm{0}$ sets a baseline for the scaling, because the 
matter density is non-vanishing even if there is no saturation.
The parameter $a$ is a scaling factor, which reflects the overall
amplitude of the density fluctuation, and the exponent $b$ is more or
less determined by the nature of hierarchical clustering. The width $w$
is in units of $h^{-1}$Mpc. Table 
\ref{tab:scaling} lists the values for LCDM model at $z = 3$ and $4.5$.
\begin{table}
\caption{Parameters in the fitting $\bar{\rho}=\rho_\mathrm{0} + a \  
(w/h^{-1}\mbox{Mpc})^b$
for the LCDM model.}
\label{tab:scaling}
\begin{tabular}{ccccc}  
\hline
$z$ & $\eta$ & $\rho_\mathrm{0}$ & a & b \\
\hline
   3    & 0.03 & 1.93  & 3.85 & 0.545 \\
        & 0.10 & 1.39  & 3.39 & 0.564 \\
  4.5   & 0.03 & 0.11  & 2.78 & 0.455 \\
        & 0.10 & 0.24  & 2.04 & 0.556 \\
\hline
\end{tabular}
\end{table}
The reason why $\rho_\mathrm{0}$ decreases with redshift is that the universe 
is more uniform early on, so that even low density regions have to absorb 
a substantial amount of Ly$\alpha$ flux to produce the low mean flux. 
This is possible because the neutral fraction of the IGM at $z=4.5$ is 
higher than that at $z=3$. As the universe evolves, the density fluctuation 
grows stronger and stronger, and the scaling factor $a$ becomes larger 
and larger. The exponent
$b$ has changed little over $z = 3$--$4.5$. The parameters also show a 
dependence on the threshold flux $\eta$, because $\eta$ sets the
threshold density $\rho_\eta$, above which the $\bar{\rho}$--$w$ relation 
is explored.

\begin{figure}  
\includegraphics[width=84mm]{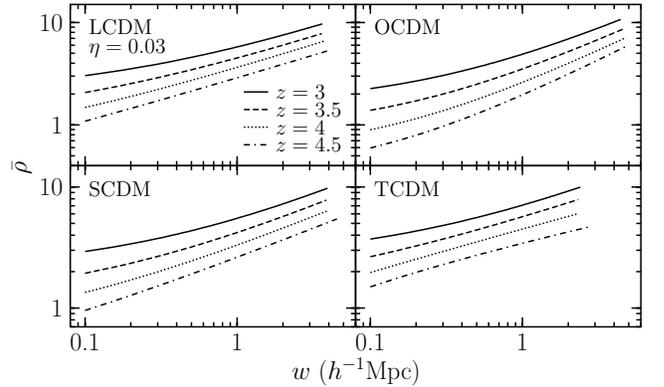}
\caption[f3]{The best-fittings of the $\bar{\rho}$--$w$ relation
for the 4 models at $z=3$--$4.5$. The threshold $\eta = 0.03$. 
The parameters are obtained by fitting the saturations with 
$w \leq 2\ h^{-1}$Mpc.}
\label{fig:scaling-lost}
\end{figure}

Fig.~\ref{fig:scaling-lost} shows the evolution of the 
$\bar{\rho}$--$w$ relation for the 4 models. The difference between models
is mostly due to the amplitude of fluctuations -- in other words, the power
spectrum.

\subsection{The Physics}
The $\bar{\rho}$--$w$ relation is analogous -- 
but not completely equivalent -- to the 
curve of growth \citep[e.g.][]{pr93}, which studies the correlation between 
the \mbox{H\,{\sc i}} column density $N_{\rm HI}$ and the 
equivalent width $W$ of Ly$\alpha$ absorptions. The similarity is as 
follows. The width $w$ is approximately the same as $W$, 
and the mean 
matter density $\bar{\rho}$ is proportional to 
$w^{-1}N_{\rm HI}$ if $\rho \propto \rho_\mathrm{b}$ holds true. Since
$N_{\rm HI} \propto W$ at small values of $W$, it is not 
surprising to see $\bar{\rho}$ grow slowly, i.e.
$\mathrm{d}\ln \bar{\rho} /\mathrm{d} \ln w \sim 0$, for small $w$.

On the other hand, the $\bar{\rho}$--$w$ relation addresses saturated
absorptions where, according to the curve-of-growth analysis, $W$ is almost 
a constant independent of the \mbox{H\,{\sc i}} column density. 
Therefore, $\bar{\rho}$ would have risen steeply against $w$, i.e. 
$b \gg 1$. This 
apparent inconsistency arises from the cosmological context, because the 
width $w$ is determined not only by $N_{\rm HI}$ but also by the physical
extent of the (relatively) dense region through Hubble expansion and 
peculiar velocities. 

Although it is not obvious why $b$ is less than unity, we can make 
an order-of-magnitude estimate using the density profile 
$\rho(r) \propto r^{-1}$ 
from a spherical self-similar infall. We modify the profile to avoid the
singularity at $r=0$ by adding a smoothing length $\epsilon$, so that 
$\rho(r) \propto (r^2+\epsilon^2)^{-1/2}$. The mean density within the 
radius $R$, at which $\rho=\rho_\eta$, is 
\begin{eqnarray} \label{eq:rho-R}
\bar{\rho}(R) &=& \frac{3}{R^3}\int_0^R \rho(r) r^2dr = 
\frac{3\rho_\eta\sqrt{R^2+\epsilon^2}}{2R^3} \times \\ \nonumber
& & \left[R\sqrt{R^2+\epsilon^2}
+\epsilon^2\log \left(\frac{\epsilon}{R+\sqrt{R^2+\epsilon^2}}\right)\right].
\end{eqnarray}
From equation (\ref{eq:rho-R}), one gets an estimate 
$b\sim \mathrm{d}\ln\bar{\rho}/\mathrm{d}\ln R = 0.39,\ 0.42$, and $0.43$ for 
$\epsilon=0.2,\ 0.3$, and $0.4\ h^{-1}$Mpc respectively at $R=2\ h^{-1}$Mpc. 
The same quantity for the $\bar{\rho}$--$w$ relation is 
$\mathrm{d}\ln \bar{\rho} /\mathrm{d} \ln w=0.41$, where we have used the 
parameters of the
LCDM model at $z=3$ with $\eta = 0.03$ (see Table \ref{tab:scaling}). It 
should be stressed that the $\rho(r) \propto r^{-1}$ profile is not quite 
justified for the IGM at $z = 3$, and without any modification, it gives the 
same mean density within the boundary $\rho=\rho_\eta$ regardless the size 
of the system, i.e. $b=0$. Therefore, equation (\ref{eq:rho-R}) is not 
expected to give a fair approximation of the $\bar{\rho}$--$w$ relation.

\section{Inversion with A Gaussian Density Profile}  \label{sec:test}

If the flux level is above the threshold, equation (\ref{eq:flux-dens}) 
can be used to find the density. When it is below the 
threshold, one must provide a density profile that matches $\bar{\rho}(w)$
and $\rho_\eta$ to fill in the missing information in the saturation. 

While the density profile is worth studying in its own right, we simply 
choose the Gaussian profile 
\begin{equation}
\rho(s) = \frac{B}{\sigma \sqrt{2\pi}}
	\exp\left[-\frac{1}{2}\bigl(\frac{s-s_\mathrm{0}}{\sigma}\bigr)^2\right],
\end{equation}
where $s$ is the coordinate in redshift space, $s_\mathrm{0}$ is the centre of 
the saturation, and $B$ and $\sigma$ are solved simultaneously from
\begin{eqnarray}
\bar{\rho}(w) &=& \frac{1}{w} \int^{s_\mathrm{0}+w/2}_{s_0-w/2} \rho(s) ds = 
\frac{B}{w}\mbox{ erf}\Bigl(\frac{1}{2\sqrt{2}}\frac{w}{\sigma}\Bigr), \\
\rho_\eta &=& \rho(s_\mathrm{0} \pm w/2) =  \frac{B}{\sigma \sqrt{2\pi}}
	\exp\left[-\frac{1}{8}\bigl(\frac{w}{\sigma}\bigr)^2\right], \nonumber
\end{eqnarray}
where erf($x$) is the error function.
The Gaussian profile has the advantage that it does
not introduce any artificial power on small scales. However, it is
arguable that the right amount of small-scale power should be added
through the profile, and so a more realistic profile may be needed. 

The recovered LOS densities with $\eta = 0.03$ and $\eta = 0.10$ are shown 
along with the four original LOS densities and fluxes in 
Fig.~\ref{fig:sample}. Since a universal density profile is 
employed in the inversion, the recovered densities do not necessarily
match the original densities.

\begin{figure}  
\includegraphics[width=75mm]{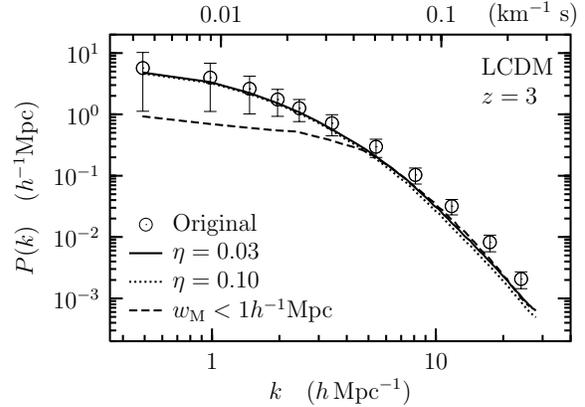}
\caption[f4]{The power spectra of original densities and 
recovered densities in redshift space. 
Circles include all LOS densities from the N-body simulation, and the 
dashed line contains only the ones that have a maximum width of 
saturation $w_\mathrm{M} < 1 \ h^{-1}$Mpc. 
The recovered densities are inverted from fluxes with thresholds 
$\eta = 0.03$ (solid line) and $\eta = 0.1$ (dotted line).
The error bars are 1$\sigma$ dispersions 
among 610 groups, each of which consists of 20 LOS densities. The error 
bars of the recovered densities, which are not plotted, are comparable to 
that of the original densities. 
\label{fig:powerspect}}
\end{figure}

\begin{figure}  
\includegraphics[width=72mm]{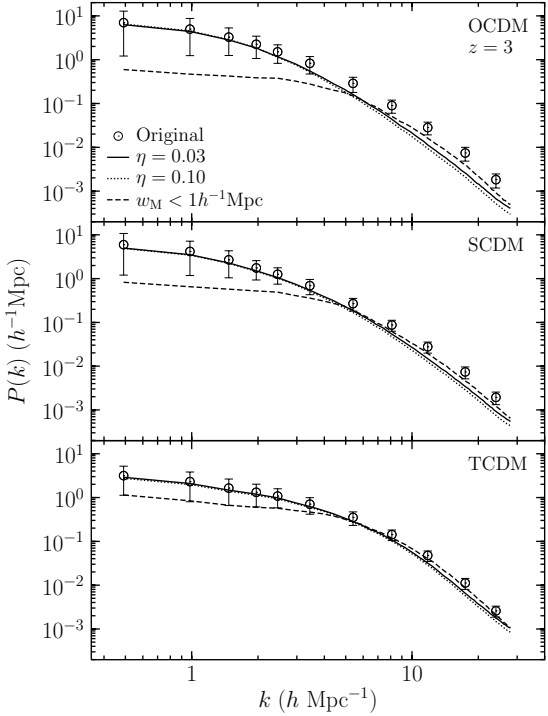}
\caption[f5]{The ame as Fig.~\ref{fig:powerspect}, but for OCDM, 
SCDM, and TCDM models.
\label{fig:powerspect-ost}}
\end{figure}

To assess the statistical quality of the inversion, we plot in 
Fig.~\ref{fig:powerspect} the original and the recovered mass power spectra of 
the LCDM model with different flux thresholds. Other models are shown 
in Fig.~\ref{fig:powerspect-ost}. The recovered power 
spectrum agrees well with the original power spectrum on large scales 
($k \la 3\ h$\,Mpc$^{-1}$), but it is underestimated on smaller scales, 
where the Gaussian profile essentially has 
no power. The signal to noise ratio, or the threshold flux, has little 
influence on large scales, but a low noise level does slightly improve the 
recovery on small scales. 

For comparison,
we show in Figs \ref{fig:powerspect} and \ref{fig:powerspect-ost} the 
original power spectrum of the LOS densities that have a maximum width of 
saturation $w_\mathrm{M} < 1 \ h^{-1}$Mpc. This is equivalent to removing -- or giving 
less weight to -- the saturated regions when one measures the power spectrum.
It is seen that the saturated regions are very important 
to scales $\ga 2\ h^{-1}$Mpc, while the unsaturated regions give a good
estimate of the small-scale power. Thus one may improve the recovery of 
power spectrum as follows. First, invert the Ly$\alpha$ forest with 
equation (\ref{eq:flux-dens}) and the $\bar{\rho}$--$w$ relation. Second, 
do the inversion after removing the saturated regions. Finally, the 
best-estimate of the power spectrum is just 
the common envelope of the power spectra of the two recovered densities. 

\begin{figure}  
\includegraphics[width=75mm]{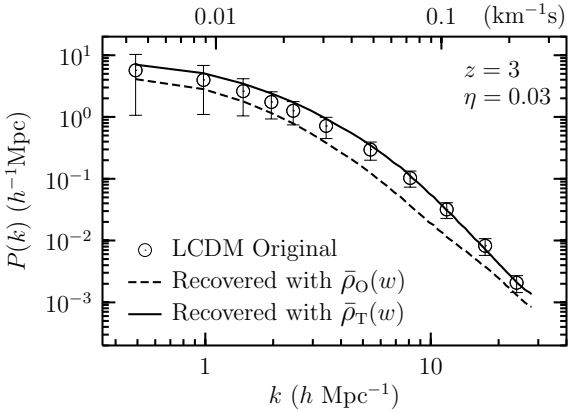}
\caption[f6]{The same as Fig.~\ref{fig:powerspect}, except that the 
fluxes are drawn from the LCDM simulation, while the densities are 
recovered with $\bar{\rho}$--$w$ relations from OCDM and TCDM simulations.
\label{fig:powerspect-moc}}
\end{figure}

Fig.~\ref{fig:scaling-lost} indicates that the $\bar{\rho}$--$w$ 
relation varies from model to model. Thus it is necessary to check if
the recovery is sensitive to $\bar{\rho}(w)$ -- in other words, if it is
model dependent. In Fig.~\ref{fig:powerspect-moc} 
we plot the power spectra of LCDM densities 
recovered with $\bar{\rho}_\mathrm{O}(w)$ and $\bar{\rho}_\mathrm{T}(w)$ from 
open cold dark matter (OCDM) and tilted cold dark matter (TCDM) 
models respectively. It seems that by boosting $\bar{\rho}(w)$ a small
amount [$\bar{\rho}_\mathrm{T}(w)$ lies a little higher than $\bar{\rho}_\mathrm{L}(w)$], one 
gets even better estimate of the power spectrum on small scales. However,
lowering $\bar{\rho}(w)$ could underestimate the power by a factor of 2
on large scales, and even more on small scales. The effect of this model 
dependence could be reduced with the constraint on small scales, since it
is possible to 
recover the small-scale power well by removing the saturations.

\begin{figure}  
\includegraphics[width=73mm]{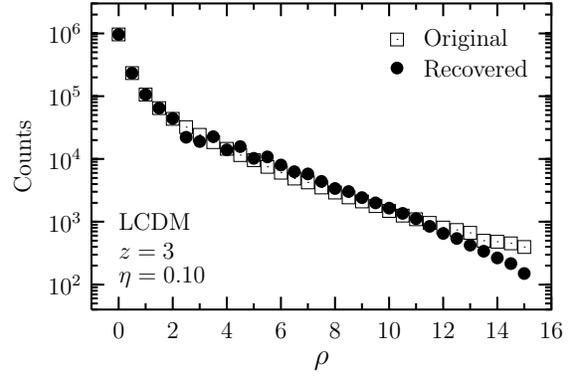}
\caption[f7]{The one-point distribution 
function of the recovered densities (filled circles) and that of the 
original densities (open squares).
\label{fig:one-point}}
\end{figure}

Fig.~\ref{fig:one-point} tests a different statistics -- the one-point 
distribution function for the LCDM model. It is evident that the Gaussian 
profile 
leads to a drop of the probability at high densities. In other words, 
it statistically reduces the heights of density peaks. 

\section{Mapping the Power Spectra}
\label{sec:map}
It seems that the inversion is no longer needed at least for determining
the mass power spectrum if one establishes a direct mapping between  
the flux and the mass power spectra \citep{c02}. It is shown that given
a set of cosmological parameters, there is a statistical mapping, which 
reliably recovers the mass power spectrum from the flux power spectrum
\citep{gh02}, even though the mapping is model dependent.

\begin{figure}  
\includegraphics[width=73mm]{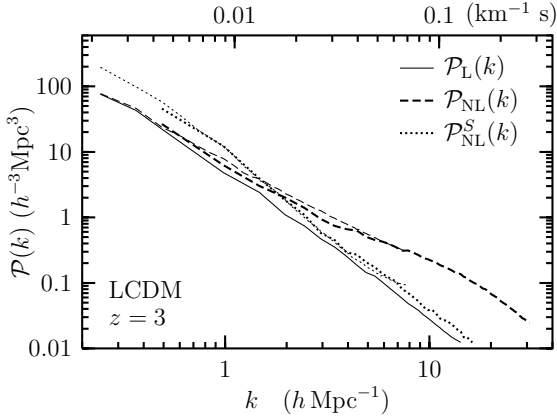}
\caption[f8]{The 3D power spectrum of the LCDM model at $z=3$. The
dashed lines are the real-space power spectra, the dotted lines are
the redshift-space power spectra averaged over solid angles, 
and the thin solid line is the 
initial power spectrum (at $z=15$) evolved to $z=3$ by linear theory. 
The thin dashed line and the thin dotted line are from an additional 
simulation with the same parameters as the LCDM model except that the 
box is 51.2 \mbox{$h^{-1}$Mpc} each side.
\label{fig:nonl}}
\end{figure}

The physical links between the three-dimensional linear mass power 
spectrum $\mathcal{P}_\mathrm{L}(k)$ and the flux power spectrum can be 
summarised by the flowchart:
\[
\mathcal{P}_\mathrm{L}(k) \longrightarrow \mathcal{P}_\mathrm{NL}(k) 
\longrightarrow \mathcal{P}_\mathrm{NL}^S(\mathbf{k}) 
\longrightarrow P(k)  \buildrel d^2(k)\over
\longrightarrow P_\mathrm{F}(k),
\]
where $\mathcal{P}_\mathrm{NL}(k)$ is the 3D non-linear mass power spectrum 
\citep{pd96, th96}, $\mathcal{P}_\mathrm{NL}^S(\mathbf{k})$ is 
$\mathcal{P}_\mathrm{NL}(k)$ 
in redshift-space \citep{k87, p99, zf03}, and 
$d^2(k) = P_\mathrm{F}(k) / P(k)$. 
Notice that $\mathcal{P}_\mathrm{NL}^S(\mathbf{k})$ is anisotropic.
We show in Fig.~\ref{fig:nonl} 
that the non-linear evolution and the redshift distortion are 
significant at $z=3$. The fact that the departure from linear evolution 
is very similar in both large and small 
box simulations indicates that such departure is real, and the cosmic 
density field has already gone non-linear below 10 \mbox{$h^{-1}$Mpc} at 
$z=3$ \citep[see also][]{zjf01,pvr01}. The angularly averaged 3D power 
spectrum $\mathcal{P}_\mathrm{NL}^S(k)$ does not give a complete view of the
difference between the real-space power spectrum and redshift-space power
spectrum, but it does show that peculiar velocities boost the power on 
large scales and reduce it substantially on small scales. 
A two-dimensional projection of $\mathcal{P}_\mathrm{NL}^S(\mathbf{k})$ can be 
found in \citet{p01}.

In contrast, the mapping in \citet{c02} follows a simplified path:
\[
\mathcal{P}_\mathrm{L}(k) \longrightarrow P_\mathrm{L}(k) \buildrel b^2(k)\over 
\longrightarrow P_\mathrm{F}(k),
\]
where $P_\mathrm{L}(k)=(2\pi)^{-1}\int_k^\infty \mathcal{P}_\mathrm{L}(y) y dy$ 
\citep{kp91}, and $b^2(k)= P_\mathrm{F}(k)/P_\mathrm{L}(k)$. Since 
$\mathcal{P}_\mathrm{NL}^S(\mathbf{k})$ is anisotropic, $P(k)$ and 
$\mathcal{P}_\mathrm{NL}^S(\mathbf{k})$ do not follow the same 
integral. Given a one-dimensional density field in redshift space 
at $z=3$, one can only
measure $P(k)$, and so $P_\mathrm{L}(k)$ is not observable unless the density
field is linear on all scales and no peculiar velocity is present. In
other words, $P(k)$ is directly connected to $P_\mathrm{F}(k)$, while $P_\mathrm{L}(k)$ 
is not.

Around the cosmic mean matter density, i.e. $|\delta| \ll 1$, equation 
(\ref{eq:flux-dens}) is approximately
\begin{equation} \label{eq:delta=0}
F(s) \simeq e^{-A}-A\beta e^{-A}\delta(s),
\end{equation}
where we have included the dependence on the redshift coordinate $s$ 
explicitly. It is seen in the linearized equation (\ref{eq:delta=0}) that 
the Fourier modes of the flux are proportional to those of the overdensity 
when $k \neq 0$. Hence, the flux power spectrum is proportional to the 
mass power spectrum
\begin{equation}
P_\mathrm{F}(k) \simeq A^2\beta^2e^{-2A}P(k), \quad k \neq 0.
\end{equation}
Caution should be taken when using equation (\ref{eq:delta=0}), because 
it is valid only if one smoothes the Ly$\alpha$ forest over large scales
\citep[see, for example,][]{zjf01, zf02}.

It is obvious from equation (\ref{eq:flux-dens}) and 
Fig.~\ref{fig:sample} that the condition $|\delta(s)| \ll 1$ is never met
where $F\simeq 1$ or $F\simeq 0$. Therefore, one has to include 
higher order terms of $\delta(s)$. Taking the simplest case, in which 
$\delta(s) = \sin(k_\mathrm{0}s)$ and only $\delta^2(s)$ term is added to 
equation (\ref{eq:delta=0}), one immediately finds that $P_\mathrm{F}(k)$ 
contains spurious power on the mode $2k_\mathrm{0}$ that $P(k)$ does not 
contain. In general, the 
non-linear density--flux relation distorts, spreads, and mixes power in 
the cosmic density field over different scales in 
the flux power spectrum. Other sources of distortion originate from 
linear filtering \citep{gh98}, line profile \citep[see][]{g92}, and 
instrumentation. We do not focus on these effects, since they are well 
studied in the literature.

\begin{figure}  
\includegraphics[width=73mm]{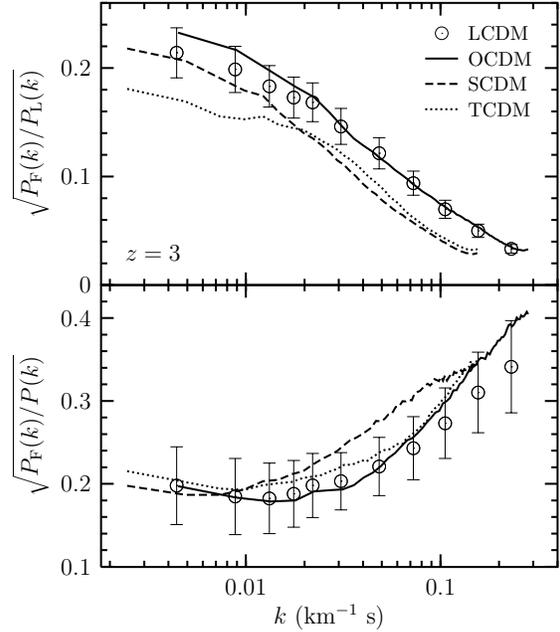}
\caption[f9]{The ratio $b(k)=\sqrt{P_\mathrm{F}(k)/P_\mathrm{L}(k)}$ (upper panel)
and $d(k)=\sqrt{P_\mathrm{F}(k)/P(k)}$ (lower panel). The Ly$\alpha$ forests 
are divided into groups, each of which consists of 20 LOS samples. 
The power spectra are averaged within each group except $P_\mathrm{L}(k)$, which 
is theoretical. The error bars of the LCDM model are 1$\sigma$ 
dispersions among 610 groups. Error bars of 
other models are comparable to that of the LCDM model. 
\label{fig:bdk}}
\end{figure}

Fig.~\ref{fig:bdk} shows the ratio $b(k)$ and $d(k)$.
The dispersion in $d(k)$ is larger than that of $b(k)$, because both
$P_\mathrm{F}(k)$ and $P(k)$ contribute to the dispersion of $d(k)$, while $b(k)$
receives only a single contribution from $P_\mathrm{F}(k)$. Even so, the error 
propagation would cause a large scatter in determining $P_\mathrm{L}(k)$. In fact, 
the scatter can be large enough that one would not be able to determine the 
cosmological model based only on the recovered $P_\mathrm{L}(k)$. On the other hand, 
the model dependence of $b(k)$ is strong enough, so that 
one may have to assume \emph{a priori} the cosmological model to recover
the mass power spectrum from the flux power spectrum. The difference among
the models is not solely due to $\sigma_\mathrm{8}$. 
For example, the standard cold dark matter (SCDM) model has a slightly 
higher $\sigma_\mathrm{8}$ than the TCDM 
model, so one would expect $b(k)$ of the SCDM model to have the same shape 
as that of the TCDM model but with a shift. This is not observed in 
Fig.~\ref{fig:bdk}, however. It is also evident from the behaviour of $b(k)$ 
and 
$d(k)$ that the power spectrum of the underlying density field $P(k)$ is 
significantly lower than the linear mass power spectrum $P_\mathrm{L}(k)$ on scales
$k>0.02$ \mbox{(km s$^{-1}$)$^{-1}$}. This is due mostly to the peculiar 
velocity, or redshift distortion.

\section{Discussion and Conclusions}  \label{sec:conclusion}
We have provided a scaling relation between the mean matter 
density and the width of the saturated Ly$\alpha$ absorptions.
It has not been tested against variations of $A$ 
or $\beta$ [see equation (\ref{eq:flux-dens})], because 
the $\bar{\rho}$--$w$ relation is the property of the density field, 
while changing $A$ or $\beta$ can only affect the width of the 
saturation, and consequently determines which part of the scaling to
be measured. This is also why the $\bar{\rho}$--$w$ relation slightly 
depends on the 
threshold flux, below which the underlying density cannot be directly 
extracted from the flux.

The inversion with the $\bar{\rho}$--$w$ relation is able to recover 
the mass power spectrum fairly well on scales above $2\ h^{-1}$Mpc, but 
underestimates the power on small scales due to the use of the Gaussian 
profile. An improvement is suggested based on
the observation that the power spectrum, with saturations removed, 
recovers the small-scale power well. Thus, by combining both the 
$\bar{\rho}$--$w$ inversion and the result after removing the saturations, 
one could get a good estimate of the power spectrum on all scales.

The Gaussian profile does not preserve the high-density tail of the 
one-point distribution function either. A more realistic 
density profile may improve the one-point distribution function, but 
the complication of substructures and peculiar velocities could result
in only a small margin for improvement.

The $\bar{\rho}$--$w$ relation provides an important constraint to 
the inversion of the Ly$\alpha$ forest. To incorporate it in any 
inversion scheme, one needs to determine the
$\bar{\rho}$--$w$ relation according to spectral 
resolution and noise. The threshold flux can be conveniently set 
to a small value above the noise level.
That is why we have not added any noise in the simulations: noise 
outside the saturation can be routinely dealt with, while noise
in the saturation, being below the threshold, does not matter.

The density 
is so far recovered in redshift space. The real-space density can be 
obtained iteratively \citep{nh99}, and the real-space three-dimensional
mass power spectrum can be recovered using matrix method \citep{h99},
or less rigorously by differentiation \citep{c98}. 

It is discussed in Section \ref{sec:sim} that the pseudo-hydro technique is 
reliable above 200 \mbox{$h^{-1}$kpc} at $z=3$. This means that while the 
bulk of our analysis is valid, the details below 200 \mbox{$h^{-1}$kpc} 
may be subject to hydrodynamic effects. In the future, we will put the 
$\bar{\rho}$--$w$ relation on test using hydrodynamical simulations. 

Finally we find that the non-linear evolution and redshift distortion
have significantly changed the mass power spectrum at $z=3$. 
Therefore, the linear mass power spectrum $P_\mathrm{L}(k)$ is not directly 
observable, and it is substantially larger than the mass power
spectrum $P(k)$ on scales $k>0.02$ \mbox{(km s$^{-1}$)$^{-1}$} at $z=3$.

\section*{acknowledgments}
HZ thanks D. Burstein, R. Dav\'e, D. Eisenstein, and L.Z. Fang for helpful 
discussions. HZ also thanks the referee for comments that help to 
improve the paper.


\begin{thebibliography}{}

\bibitem[\protect\citeauthoryear{Aguirre, Schaye \& Theuns}{Aguirre et al.}
{2002}]{ast02} Aguirre A., Schaye J., Theuns T., 2002, ApJ, 576, 1

\bibitem[\protect\citeauthoryear{Bardeen et al.}{1986}]{b86} Bardeen J.M., 
Bond J.R., Kaiser N., Szalay A.S., 1986, ApJ, 304, 15

\bibitem[\protect\citeauthoryear{Bi}{1993}]{b93} Bi H.G., 1993, ApJ, 405, 479

\bibitem[\protect\citeauthoryear{Bi \& Davidsen}{1997}]{bd97} Bi H.G., 
Davidsen A.F., 1997, ApJ, 479, 523 

\bibitem[\protect\citeauthoryear{Bi, Ge \& Fang}{Bi et al.}{1995}]{bgf95} 
Bi H.G., Ge J., Fang L.Z., 1995, ApJ, 452, 90

\bibitem[\protect\citeauthoryear{Binney \& Merrifield}{1998}]{bm98}
Binney J., Merrifield M., 1998, Galactic Astronomy, Princeton Univ. Press, 
Princeton, NJ, p. 110

\bibitem[\protect\citeauthoryear{Bryan et al.}{1999}]{b99} Bryan G.L.,
Machacek M., Anninos P., Norman M.L., 1999, ApJ, 517, 13

\bibitem[\protect\citeauthoryear{Cen et al.}{1994}]{c94} Cen R.,
Miralda-Escud\'e J., Ostriker J.P., Rauch M., 1994, ApJ, 437, L9

\bibitem[\protect\citeauthoryear{Cowie \& Songaila}{1998}]{cs98}
Cowie L.L., Songaila A., 1998, Nature, 394, 44

\bibitem[\protect\citeauthoryear{Croft et al.}{1998}]{c98} Croft R.A.C., 
Weinberg D.H., Katz N., Hernquist L., 1998, ApJ, 495, 44

\bibitem[\protect\citeauthoryear{Croft et al.}{1999}]{c99} Croft R.A.C., 
Weinberg D.H., Pettini M., Hernquist L., Katz N., 1999, ApJ, 520, 1

\bibitem[\protect\citeauthoryear{Croft et al.}{2002}]{c02} Croft R.A.C., 
Weinberg D.H., Bolte M., Burles S., Hernquist L., Katz N., Kirkman D., 
Tytler D., 2002, ApJ, 581, 20

\bibitem[\protect\citeauthoryear{Dav\'e et al.}{1999}]{dhk99} Dav\'e R.,
Hernquist L., Katz N., Weinberg D.H., 1999, ApJ, 511, 521

\bibitem[\protect\citeauthoryear{Dav\'e et al.}{2001}]{dco01} Dav\'e R.,
Cen R., Ostriker J.P., Bryan G.L., Hernquist L., Katz N., Weinberg, D.H.,
Norman M.L., O'Shea B., 2001, ApJ, 552, 473

\bibitem[\protect\citeauthoryear{Gnedin \& Hamilton}{2002}]{gh02}
Gnedin N.Y., Hamilton A.J.S., 2002, MNRAS, 334, 107

\bibitem[\protect\citeauthoryear{Gnedin \& Hui}{1998}]{gh98} Gnedin N.Y.,
Hui L., 1998, MNRAS, 296, 44

\bibitem[\protect\citeauthoryear{Gray}{1992}]{g92} Gray D.,
1992, The Observation and Analysis of Stellar Photospheres,  
Cambridge Univ. Press, New York

\bibitem[\protect\citeauthoryear{Hockney \& Eastwood}{1981}]{he81}
Hockney R.W., Eastwood J.W., 1981, Computer Simulation Using Particles.
McGraw-Hill, New York

\bibitem[\protect\citeauthoryear{Hui}{1999}]{h99} Hui L., 1999, ApJ, 516, 519

\bibitem[\protect\citeauthoryear{Hui \& Gnedin}{1997}]{hg97} Hui L.,
Gnedin N.Y., 1997, MNRAS, 292, 27

\bibitem[\protect\citeauthoryear{Jing \& Fang}{1994}]{jf94} 
Jing Y.P., Fang L.Z., 1994, ApJ, 432, 438

\bibitem[\protect\citeauthoryear{Kaiser}{1987}]{k87}
Kaiser N., 1987, MNRAS, 227, 1

\bibitem[\protect\citeauthoryear{Kaiser \& Peacock}{1991}]{kp91}
Kaiser N., Peacock, J.A., 1991, ApJ, 379, 482

\bibitem[\protect\citeauthoryear{Kim et al.}{2002}]{k02}
Kim T.-S., Carswell R.F., Cristiani S., D'Odorico S., 
Giallongo E., 2002, MNRAS, 335, 555

\bibitem[\protect\citeauthoryear{Lu et al.}{1996}]{l96} Lu L.,
Sargent W.L.W., Womble D.S., Takada-Hidai M., 1996, ApJ, 472, 509

\bibitem[\protect\citeauthoryear{McDonald et al.}{2000}]{m00} McDonald P.,
Miralda-Escud\'e J., Rauch M., Sargent W.L.W., Barlow T.A., Cen R., 
Ostriker J.P., 2000, ApJ, 543, 1

\bibitem[\protect\citeauthoryear{Navarro, Frenk \& White}{Navarro et al.}
{1996}]{nfw96} Navarro J.F., Frenk C.S., White S.D.M., 1996, ApJ 462, 563

\bibitem[\protect\citeauthoryear{Nusser \& Haehnelt}{1999}]{nh99} 
Nusser A., Haehnelt M., 1999, MNRAS, 303, 179

\bibitem[\protect\citeauthoryear{Peacock}{1999}]{p99} Peacock J.A., 
1999, Cosmological Physics, Cambridge Univ. Press, Cambridge

\bibitem[\protect\citeauthoryear{Peacock \& Dodds}{1996}]{pd96} 
Peacock J.A., Dodds S.J., 1996, MNRAS, 280, L19

\bibitem[\protect\citeauthoryear{Peacock et al.}{2001}]{p01}
Peacock J.A. et al., 2001, Nature, 410, 169

\bibitem[\protect\citeauthoryear{Petitjean, M\"ucket \& Kates}
{Petitjean et al.}{1995}]{pmk95} Petitjean P., M\"ucket J.P., 
Kates R.E., 1995, A\&A 295, L9

\bibitem[\protect\citeauthoryear{Pichon et al.}{2001}]{pvr01} Pichon C., 
Vergely J.L., Rollinde E., Colombi S., Petitjean P., 2001, MNRAS, 326, 597

\bibitem[\protect\citeauthoryear{Press \& Rybicki}{1993}]{pr93}
Press W.H., Rybicki G.B., 1993, ApJ, 418, 585

\bibitem[\protect\citeauthoryear{Rauch et al.}{1997}]{r97} Rauch M.,
Miralda-Escud\'e J., Sargent W.L., Barlow T.A., Weinberg D.H.,
Hernquist L., Katz N., Cen R., Ostriker J.P., 1997, ApJ, 489, 7

\bibitem[\protect\citeauthoryear{Riediger, Petitjean \& M\"ucket}
{Riediger et al.}{1998}]{rpm98} Riediger R., Petitjean P., 
M\"ucket J.P., 1998, A\&A, 329, 30

\bibitem[\protect\citeauthoryear{Schaye}{2001}]{s01}
Schaye J., 2001, ApJ, 559, 507

\bibitem[\protect\citeauthoryear{Taylor \& Hamilton}{1996}]{th96}
Taylor A.N., Hamilton A.J.S., 1996, MNRAS, 282, 767

\bibitem[\protect\citeauthoryear{Viel et al.}{2002}]{v02}  Viel M., 
Matarrese S., Mo H.J., Haehnelt M.G., Theuns T., 2002, MNRAS, 329, 
848

\bibitem[\protect\citeauthoryear{Zhan \& Fang}{2002}]{zf02} Zhan H.,
Fang, L.Z., 2002, ApJ, 566, 9

\bibitem[\protect\citeauthoryear{Zhan \& Fang}{2003}]{zf03} Zhan H.,
Fang, L.Z., 2003, ApJ, 585, 12

\bibitem[\protect\citeauthoryear{Zhan, Jamkhedkar \& Fang}{Zhan et al.}
{2001}]{zjf01} Zhan H., Jamkhedkar P., Fang, L.Z., 2001, ApJ, 555, 58

\bibitem[\protect\citeauthoryear{Zhang et al.}{1998}]{z98} Zhang Y.,
Meiksin A., Anninos P., Norman M.L., 1998, ApJ, 495, 63

\end{thebibliography}
\end{document}